\documentclass[12pt,preprint]{aastex}

\newcommand\Mpc{$h^{-1}$Mpc}
\newcommand\etal{{\it et al.\ }}

\newcommand\Om{{$\Omega_m$}}

\begin{document}

\title{An Inversion Method for Measuring $\beta$ in Large Redshift Surveys}
\author{Stephen D. Landy}
\affil{Department of Physics, College of William and Mary,
     Williamsburg, VA 23187-8795, USA}
\email{landy@physics.wm.edu}
\and
\author{Alexander S. Szalay}
\affil{Department of Physics and Astronomy, The Johns Hopkins University,
Baltimore, MD 21218, USA}

\begin{abstract}

A precision method for determining the value of $\beta=\Omega_m^{0.6}/b$, where
$b$ is the galaxy bias parameter, is presented. In contrast to other existing
techniques that focus on estimating this quantity by measuring distortions in
the redshift space galaxy-galaxy correlation function or power spectrum, this
method removes the distortions by reconstructing the real space density field
and determining the value of $\beta$ that results in a symmetric signal. To
remove the distortions, the method modifies the amplitudes of a Fourier
plane-wave expansion of the survey data parameterized by $\beta$. This
technique is not dependent on the small-angle/plane-parallel approximation and
can make full use of large redshift survey data. It has been tested using
simulations with four different cosmologies and returns the value of $\beta$ to
$\pm0.031$, over a factor of two improvement over existing techniques.

\end{abstract}


\section{Introduction}

One of the principal goals of the current large redshift surveys, such as the
Two-Degree Field (2dF) and Sloan Digital Sky Surveys (SDSS), is an improved
estimation of the total mass density parameter $\Omega_m$. A highly accurate
measurement of \Om, along with measurements of the acoustic peaks in the CMB
and the deceleration parameter $q_o$ from distant supernovae, should enable
cosmologists to pin down the global geometry of the Universe and infer the
value of $\Omega_{\lambda}$, the vacuum energy density.

As is well-known, the redshift of a galaxy represents the sum of the galaxy's
redshift distance, which depends upon cosmology, and its radial peculiar
velocity. Therefore, raw redshift distance measurements are contaminated with
the galaxies' radial peculiar velocities. Consequently, galaxy redshift survey
statistical measures such as the two-point galaxy correlation function contain
distortions due to the existence of galaxy peculiar velocities.

On large-scales in the linear regime, the galaxy peculiar velocities can be
identified with fractional perturbations in the Hubble ratio $\Delta H/H$ due
to fluctuations in the matter density field $\Delta\rho/\rho$, with $\Delta H/H
\propto \Omega_m^{0.6}\Delta\rho/\rho$.\footnote{Assuming $\Omega_\lambda = 0$.
Otherwise $\Omega_m^{0.6} \rightarrow \Omega_m^{0.6} + {\Omega_\lambda\over
70}(1+{\Omega_m\over2})$ (see Lahav \etal 1991)} The peculiar velocities and
resultant redshift distortions are statistical and by utilizing measures of the
galaxy mass field, such as the galaxy-galaxy correlation function, together
with the relation above, the distortions can be modeled and exploited as a way
to measure $\Omega_m$.

Since $\Omega_m$ is estimated using distortions in the galaxy correlation
function or power spectrum and it is believed that galaxy formation is biased
with respect to the underlying total mass field, what is actually measured is
the parameter $\beta=\Omega_m^{0.6}/b$, where $b$ is the bias in galaxy
clustering with respect to the underlying mass field. Although somewhat
problematic, this holds for any method that relies on the clustering of
luminous matter to estimate a dynamical measure of the total mass field.

\subsection{Kaiser's Method}

The utility of using distortions in the galaxy redshift space correlation
function to measure $\beta$ in the linear regime was developed in detail by
Kaiser (1987), although its use was anticipated by others (see Sargent \&
Turner 1977; Peebles 1980 \S 76). In this seminal work, Kaiser's unique
contribution was in showing that the anisotropies generated by these radial
peculiar velocities could be modeled as redshift space density enhancements in
the underlying mass field.

Following Kaiser's lead, let us take a plane wave density fluctuation
\begin{equation}
\Delta_{r}({\bf r}) = \Delta_k\cos({\bf k\cdot r + \theta})
\end{equation}
in real space. This real space plane wave fluctuation maps to redshift space in
the following simple way,

\begin{equation}
\Delta_{s}({\bf r}) = \Delta_r({\bf r}) [1+{\Omega_m^{0.6}\over b}\cos({\bf
\hat{k}\cdot \hat{r})}^2]
\end{equation}
or
\begin{equation}
\Delta_{s}({\bf r}) = \Delta_r({\bf r}) (1+\beta \mu^2)
\end{equation}
where $\mu=\cos({\bf \hat{k}\cdot \hat{r}})$ is the cosine of the angle between
the normed wavevectors ${\bf \hat{k}}$ and the line-of-sight ${\bf \hat{r}}$,
$\Delta_{s}$ is the redshift space density, and $\beta=\Omega_m^{0.6}/b$. In
the limit of small angular separations, the distortions in the redshift space
power spectrum $P_s({\bf k})$ are given by
\begin{equation}
P_s({\bf k}) = P_r({\bf k})(1+\beta\mu^2)^2
\end{equation}
where $P_r({\bf k})$ is the real space power spectrum. This limit is also known
as the plane-parallel or distant observer approximation. Since the distortions
contain powers of $\mu^2$ and $\mu^4$, they appear as a combination of
quadrupole and hexadecapole distortions in $P_s({\bf k})$. As Kaiser went on to
note, it seemed more practical to find a way to determine $\beta$ by using this
result to calculate distortions in the redshift space correlation function, the
correlation function being the Fourier transform of the power spectrum.

\section{Difficulties in Measuring the Distortions}

Traditionally, the approach to measuring distortions in the correlation
function has been to decompose it in two-dimensions, where the two axes
correspond to the directions parallel ($\pi$) and perpendicular ($r_{p}$) to
the line-of-sight to a pair of galaxies. The line-of-sight direction has been
taken to be the direction of the center-of-mass (Peebles 1980) or the
half-angle (Landy, Szalay, \& Broadhurst 1998; Hamilton 1992) between the pair.
The resulting correlation function is denoted by $\xi_{z}(r_{p},\pi)$. This
decomposition for the correlation function for the simulation data is shown in
Figure 1 on the following page.

\begin{figure}[h]
\vbox to2.3in{\rule{0pt}{3.5in}} \includegraphics{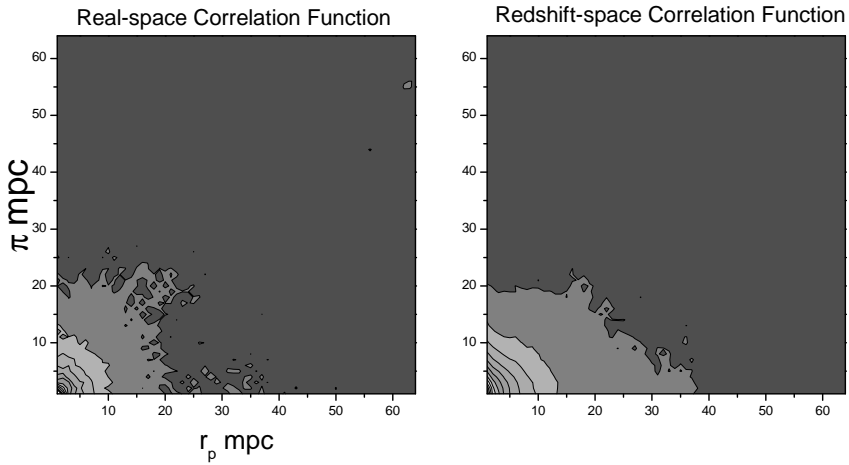} \caption{\it The real and redshift space
correlation functions from one of the simulations. The expansion in the
redshift space function at small separations along the $\pi$-axis is due to the
small-scale non-linear pairwise velocity dispersion. The compression at larger
separations is due to large-scale infall, which is a function of $\beta$.}
\end{figure}

However, measuring $\beta$ from distortions in the correlation function has
proven difficult for several reasons. Firstly, as evident in Figure 1, the
distortions at small separations are due to galaxy pairwise velocities in the
non-linear regime. These distortions are of an opposite sense than those due to
large-scale linear infall. Whereas the latter cause a compression in the
correlation function along the line-of-sight, the non-linear distortions create
an expansion. Therefore, they are competing effects and errors in their
measurement may be positively correlated when both are estimated
simultaneously. Secondly, each of these distortions affect the correlation
function in terms of convolutions. Convolutions are troublesome since it is the
unknown underlying function convolved with the kernel that is actually
estimated. And, since the distortions enter in the radial direction only,
calculations of the expected correlation function given a particular power
spectrum are extremely complicated and involve an integration over the spectrum
itself, whose value may be only weakly constrained. For example, Kaiser's
original paper derives the correlation function for the case of linear infall
for {\it just the line-of-sight direction} as:
\begin{eqnarray}
 \xi_{z}(0,r_p)=4\pi\int_{0}^{\infty} dkk^2 P_r(k)\Biggl\lbrack \sin(kr)\left\{{1\over
kr} +2\Omega^{0.6}{{[(kr)^2-2]}\over
(kr)^3}+\Omega^{1.2}{{[(kr)^4-12(kr)^2+24]}\over (kr)^5}\right\} \cr
+\cos(kr)\left\{{{4\Omega^{0.6}}\over
(kr)^2}+\Omega^{1.2}{{[4(kr)^2-24]}\over(kr)^4}\right\}\Biggr\rbrack
\end{eqnarray}
with $b=1$. And thirdly, unless measurements of the correlation function are
restricted to small angular separations for the galaxy pairs (the distant
observer approximation), the expected distortions in the correlation function
become even more complex. The complete problem in the case of wide-angles was
recently solved in Szalay, Matsubara, \& Landy (1998) and computer algebra was
needed to keep track of the numerous terms.

Attempting to measure the distortions using the power spectrum rather than the
correlation function has also proven problematic. The basic intractability of
this problem mainly results from the fact that radial distortions are not
translationally invariant. The simple form for the distortions $P_s({\bf k}) =
P_r({\bf k})(1+\beta\mu^2)^2$ given by Kaiser is only valid in the limit of
small angles. Even with today's redshift surveys, this severely limits the
amount of data that can be used in an analysis. A full decomposition of the
distortions in the power spectrum in the case of wide-angles again results in
very complex equations and requires a model for the underlying power spectrum
since the distortions cross-correlate Fourier modes on very different scales.

Many researchers have been working over the last two decades on developing
clever and robust methods to model the distortions and these methods have been
based principally upon one of three approaches. The first involves measuring
the ratio of the angle-averaged redshift space to real space power spectrum or
correlation function (see Fry \& Gazta\~naga 1994; Peacock \& Dodds 1994; Baugh
1996; Loveday \etal 1996; Tadros \& Efstathiou 1996; Peacock 1997). The second
method calculates the ratio of the quadrupole to monopole moments of the
redshift space power spectrum (see Hamilton 1993a,1995,1997a; Bromley 1994;
Fisher \etal 1994; Cole, Fisher, \& Weinberg 1994,1995; Lin 1995; Fisher \&
Nusser 1996; Taylor \& Hamilton 1996; Bromley, Warren, \& Zurek 1997). The
third utilizes a maximum likelihood approach in which the amplitudes of
individual modes, $\beta$, and the power spectrum are taken as parameters of
the model (see Fisher, Scharf, \& Lahav 1994; Heavens \& Taylor 1995;
Ballinger, Heavens, \& Taylor 1995). An excellent review of these methods along
with a thorough background of the linear redshift distortion problem is given
in Hamilton (1997a).

Other work dealing primarily with the development of new methods, theoretical
analysis, and constraints on surveys can be found in Lahav \etal (1991); Suto,
\& Suginohara (1991); Hamilton (1992,1993b,1997); Gramann, Cen, \& Bachall
(1993); Gramann, Cen, \& Gott (1994); Hamilton \& Culhane (1996); Matsubara, \&
Suto (1996); Zaroubi, \& Hoffman (1996); de Laix \& Starkman (1998); Szalay,
Matsubara, \& Landy (1998); Hatton \& Cole (1998,1999); and  Nakamura,
Matsubara, \& Suto (1998).

In more recent work, Tadros \etal (1999) present a spherical harmonic analysis
of the linear distortions in the PSCz Galaxy Catalog and include a detailed
synopsis of many of the linear distortion measurements listed above. They
measure $\beta=0.58\pm0.26$. In Matsubara, Szalay, \& Landy (2000), the first
application of the Karhunen-Loève (K-L) eigenmode compression technique to this
problem is presented using the Las Campanas Redshift Survey data with an
estimation of $\beta=0.30\pm 0.39.$ Hamilton, Tegmark, \& Padmanabhan (2000)
analyze the PSCz catalog also using a K-L expansion technique and determine
$\beta=0.41_{-0.12}^{+0.13}$. A modified approach incorporating eigenvectors is
described in Taylor \etal (2001) who find $\beta=0.39{_{-0.12}^{+0.14}}$.
Ratcliffe \etal (1998) using the Durham/UKST Galaxy Redshift Survey estimate
$\beta\sim 0.5$. Peacock \etal (2001), using over 141,000 galaxies from the
Two-Degree Field Survey report a value of  $\beta=0.43\pm0.07$ using a two
parameter model of the distortions in the redshift space correlation function.
Tegmark, Hamilton, \& Xu (2000) apply a pseudo K-L eigenmode analysis to the
Two-Degree Field public release data and estimate $\beta=0.49\pm0.16$.

\section{Removing the Distortions}

All of the methods listed above depend upon a measurement of the distortion
signal, subject to all its complexities. However, Kaiser's identification of
the redshift distortions with a statistically equivalent distortion of the real
space mass density presents another approach to the problem. In Tegmark \&
Bromley (1995), an analytical reconstruction of a real space density field from
a redshift space density field, such as that derived from a redshift survey,
was described. However, as with attempts to model the distortions, the
procedure had a complicated functional form and involved a relatively high
dimensional integral. Because of this, it was suggested that such as inversion
was best handled by parallel computation on a supercomputer. Setting
computational complexities aside, the reconstruction of the real space density
field presents a powerful new approach to this problem.

As previously discussed, a direct fit of a model to either the correlation
function or power spectrum in redshift space presents numerous difficulties due
to the fundamental complexity of the effects of the distortions. Additionally,
many of these methods include assumptions concerning either the underlying
power spectrum or correlation function on large linear scales. However, rather
than measuring the distortions themselves, another approach would be to invert
the density enhancements and statistically re-create the real space density
field. Since two-point estimators such as the power spectrum and correlation
function are statistical anyway, one could invert the density field and simply
search for the value of $\beta$ that results in a symmetric function. In other
words, when the field is properly inverted the anisotropies disappear.

This method would work as follows:\\ \\ \indent 1) Take the data and perform
the inversion for different values of the redshift distortions\\ \indent  as
parameterized by $\beta$.\\ \indent 2) For each of these inversions, calculate
the power spectrum or correlation function \\ \indent that results from the
inverted data.\\  \indent 3) Look for the appropriate symmetry in the resulting
function.\\
\\ Here the problem is reduced from the having to calculate and fit a very
complex function with an assumed power spectrum, to applying an inversion
technique to the data itself and looking for a symmetric signal.

\subsection{A Novel Method for Removing the Distortions}

Consider again Kaiser's analysis of the redshift space distortions. The power
of his approach resulted from considering a plane wave density fluctuation in
real space. As shown above, this fluctuation maps to redshift space as
\begin{equation}
\Delta_{s}({\bf r}) = \Delta_r({\bf r}) (1+\beta\mu^2)
\end{equation}
where ${\bf r}$ the line-of-sight in either real or redshift space. This
relation is easily inverted to first order to give the real space density
fluctuation,
\begin{equation}
\Delta_{r}({\bf r}) = {{\Delta_s({\bf r})}\over {1+\beta\mu^2}}.
\end{equation}

At first this development may seem academic in that redshift surveys do not
appear to consist of plane wave density fluctuations. However, a redshift
survey consists of a collection of galaxies, which for all intents and purposes
can be considered point particles. As point particles, it is simple to
construct a $\delta$-function plane wave expansion of each individual point.
Since each point has a well-defined position vector, it is easy to calculate
the {\it undistortion} coefficient $1/(1+\beta\mu^2)$ for each wavevector in
the expansion for a given $\beta$. Once this is done, the real space density
can be reconstructed by re-summing this same set of plane waves with their
modified amplitudes.

Mathematically, the procedure is very straightforward. The real space mass
density $\rho({\bf r})$ at any point ${\bf r}$ is given by
\begin{equation}
\rho({\bf r}) =   {1 \over n} \sum_{i=1}^{Ngal} \sum_{j=1}^{n} {\left( 1 \over
{1+\beta\mu_{ij}^2}\right)} {\rm{exp}[i({\bf k_j\cdot r}-{\bf k_j\cdot r_i})]}.
\end{equation}
Here each galaxy $i$ at position ${\bf r_i}$ is expanded in a $\delta$-function
plane wave expansion over a set of $n$ wavevectors ${\bf k_j}$, the amplitude
of each wavevector is divided by the inversion coefficient,
${1+\beta\mu_{ij}^2}$ where $\mu_{ij}=\cos({\bf \hat{k}_j\cdot \hat{r}_i})$,
the sum taken over all galaxies and the density reconstructed at ${\bf r}$.
Given the rotational but not translational invariance of these distortions, it
is fortunate that the inversion of this problem can be expressed so simply in
Cartesian coordinates using the Fourier transform.

In practice this procedure is only slightly more complicated than that
described above. The complications have to do with accounting for survey
geometry, taking into account the non-linear small-scale pairwise distortions,
choosing a finite set of wavevectors to expand over, and developing a sensitive
method for determining when the $\beta$ distortions have been removed.

\section{Testing the Method}

To develop and test the method, numerical simulations were used (see Cole \etal
1999). Simulations are very convenient since the parameters of the `data' are
known and real space positions of the galaxies are also available. To make the
procedure computationally efficient, the analysis was restricted to thin slices
of simulated data 2$\degr$ thick by 55$\degr$ wide by 200 to 500 \Mpc~deep.
Each slice contained approximately 5000 galaxies and the radial selection
function was modeled after that of the SDSS. As an initial test, simulation
data spanning four different cosmologies were used: two Cobe normalized models
with $\beta=0.57$ ($\Omega_m=0.5$), $\beta=0.26$ ($\Omega_m=0.3$), one
tilted-flat model with $\beta=0.51$ ($\Omega_m=1.0$), and two independent
$\tau$CDM flat models with $\beta=0.54$ ($\Omega_m=1.0$). These are named O3,
O5, E2, and E3SA, E3SB in Cole \etal (1999), respectively.

Central to the method is to first expand the data in a Fourier expansion. Each
point must be expanded and the new Fourier amplitudes calculated individually
since the distortions depend upon ${\cos(\bf \hat{k}\cdot \hat{r}})$. In
computing the Fourier expansion, the data was binned with a resolution 1\Mpc~
using a grid of $1024\times1024$ in two dimensions and a FFT performed. By
using very thin slices, the expansion could be carried out in two rather than
three dimensions, limiting the computational effort. Further work will
efficiently adapt this method to three dimensions. On this scale in two
dimensions there are already $512\times1024$ amplitudes that must be
calculated.

After all amplitudes have been modified, their Fourier amplitudes are added
back together and an inverse FFT performed. This operation returns the
undistorted density field. The field is then cut to correspond to the geometry
of the original slice. This procedure was carried out for twenty slices and the
$\xi_{z}(r_{p},\pi)$ correlation function calculated for each one. All twenty
correlation functions were averaged together to construct the mean. The mean
correlation function was then windowed with a Hann window with a limit of
64\Mpc~ and Fourier transformed. This returned the power spectrum of the
$\xi_{z}(r_{p},\pi)$ decomposition.

Although the focus of this method is to measure $\beta$, as was shown in Figure
1,~$\xi_{z}(r_{p},\pi)$ also contains small-scale pairwise velocity distortions
in the non-linear regime. These distortions have been shown to be
well-characterized by an exponential distribution, which is equivalent to a
Lorentzian distortion in the power spectrum (see Landy, Szalay \& Broadhurst
1998; Landy 2002). The full distortion function of the power spectrum is given
by the ratio
\begin{equation}
{{P_s({k})}\over{P_r({k})}} = {{(1+\beta\mu^2)^2}\over{1+{1\over2}
k^2\mu^2\sigma^2_{12}(k)}}
\end{equation}
where $\sigma_{12}/\sqrt{2}$ is the decay width of the exponential. Using
simulated data, it is simple to construct this ratio by dividing the power
spectrum of the redshift space data by that of real space data. The ratio of
these power spectra is shown in Figure 2. The coordinate system is the Fourier
analogue of the $(r_{p},\pi)$ basis and is denoted by ($k_{r_p},k_{\pi})$.

\begin{figure}[h]
\vbox to2.1in{\rule{0pt}{3.5in}} \includegraphics{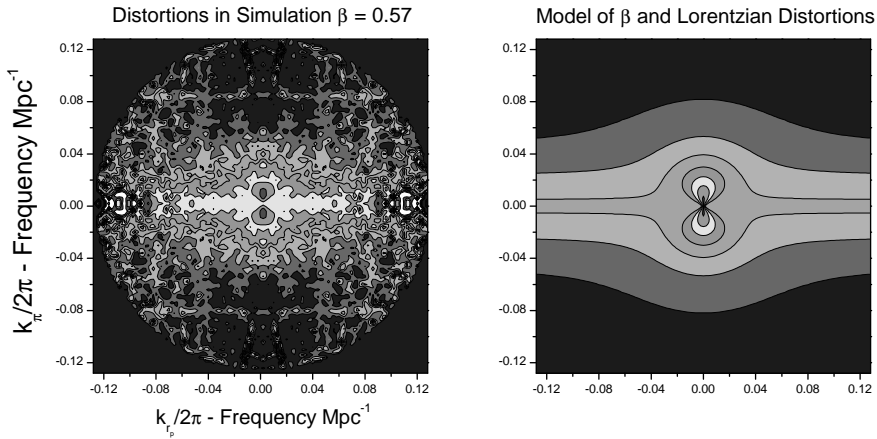} \caption{\it The left panel shows the
distortions in the power spectrum for the simulated data. This was constructed
by dividing the power spectrum in redshift space by the power spectrum in real
space, using the real and redshift space functions from the simulated data. On
the right is an analytical representation of the same function with
$\beta=0.57$ and $\sigma_{12}=360$~km/sec. The agreement between the analytical
model and the distortions is quite good. The `bubble' in the center (long
wavelengths) shows the linear infall distortions. The distortions across the
graph are due to the small-scale pairwise velocity dispersion function (the
Lorentzian).}
\end{figure}

As shown in Figure 2, most of the signal from the $\beta$ distortions is found
near the core. However, since the $\beta$ and small-scale distortions are
competing effects, systematic errors will result if the small-scale distortions
are not modeled correctly, especially at larger scales (in the core), and these
errors will be positively correlated. A method to overcome the problem of
modeling the small-scale distortions is described below.

\subsection{Removing the Distortions from Redshift Space Data}

To test the method, the procedure described above was used to remove the
$\beta$ distortions from the redshift space data. In practice, the distorted
data was smoothed on 8 mpc scales and the correlation function windowed at 64
mpc to limit noise. Figure 3 shows the original smoothed and windowed power
spectrum, and the power spectra for three values of $\beta$ used to remove the
distortions.

\begin{figure}[h]
\vbox to2.0in{\rule{0pt}{3.5in}} \includegraphics{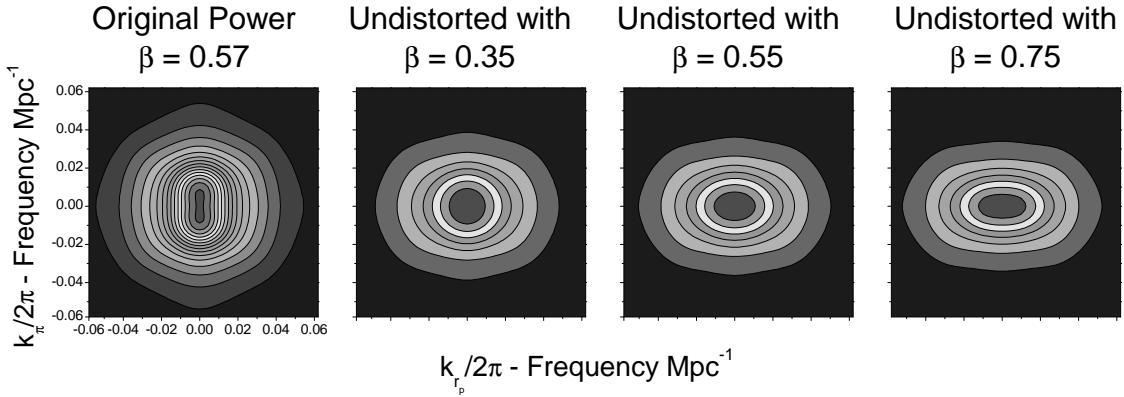} \caption{\it The original power spectrum, and
the power spectra for three sample values of $\beta$ used to remove the
distortions for the $\beta=0.57$ simulation. Although the case of $\beta=0.35$
appears to be the most symmetric result, the small scale non-linear distortions
have not yet been taken into account. As will be shown in Figure 4, the
vertically compressed signal of $\beta=0.55$ is actually very close to the
expected result.}
\end{figure}

Due to the residual non-linear distortions, the end result of this procedure is
not expected to be a circularly symmetric function. However, the non-linear
distortions effectively act only in the $k_{\pi}$ direction, that is along the
line of sight. Therefore, even when the $\beta$ distortions have been
completely removed what is expected is a vertically compressed function about
the $k_{r_p}$ axis.

What is needed at this stage is a method to account for the non-linear
distortions. One of the fundamental challenges of any method is in trying to
simultaneously fit the $\beta$ and non-linear distortions. The main problem
derives from the situation that these distortions are competing effects and
their errors are positively correlated.

To overcome this difficulty, a procedure was developed to estimate the
non-linear distortions from the data itself, taking advantage of only the
geometrical knowledge of how they effect the power spectrum. This also belies
any problem which might arise from smoothing and windowing the data before
transforming.

In theory, neither the $\beta$ nor the non-linear distortions effect the power
along the  $k_{r_p}$ axis, so the power along this axis can be utilized as the
model underlying power spectrum. Additionally, if the $\beta$ distortions have
been removed in their entirety, the only remaining signal should be due to the
residual non-linear distortions, which should be independent of $k_{r_p}$.
Figure 4 shows the power spectrum of the data divided by the expected power as
a function of $|k|$, given by the signal along the $k_{r_p}$ axis. The power is
almost perfectly symmetrical and independent of $k_{r_p}$ axis about the
$k_{r_p}$ axis near the correct value of $\beta=0.57$. Either under-correction
using $\beta=0.35$ or over-correction using $\beta=0.75$, leave obvious
signatures in the result.

\begin{figure}[h]
\vbox to2.0in{\rule{0pt}{2.5in}} \includegraphics{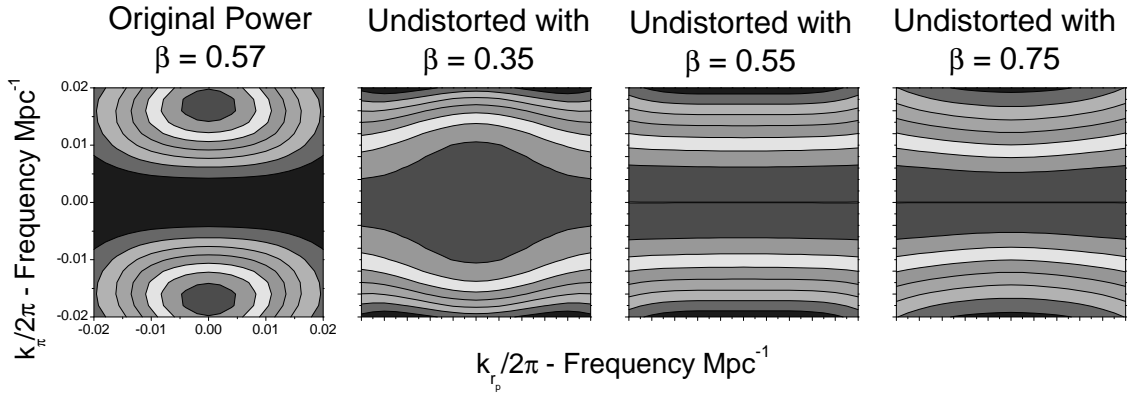} \caption{\it The original power spectrum and
that for three sample values of $\beta$ used to remove the distortions divided
by the model power spectrum as derived from the data (the value along the
$k_{r_p}$ axis). If the $\beta$ distortions have been correctly removed, this
function should be symmetric about the $k_{r_p}$ axis. The cases for
$\beta=0.35$ and $\beta=0.75$ clearly show under and over corrections.}
\end{figure}

\subsection{A Sensitive Fit}

The ratios shown in Figure 4 illustrate the utility of the technique. To
actually determine the best fit value of $\beta$, however, the numerical
difference between the intrinsic model derived from the data, that is the
signal along the $k_{r_p}$ axis, and the undistorted spectra for different
values of $\beta$ was used. From this, a generalized `chi-square' function was
calculated to determine the optimal value of $\beta$. Figure 5 shows this
difference function for several values of $\beta$ although over a much smaller
range to illustrate the sensitivity of this method.

\begin{figure}[h]
\vbox to1.7in{\rule{0pt}{3.5in}} \includegraphics{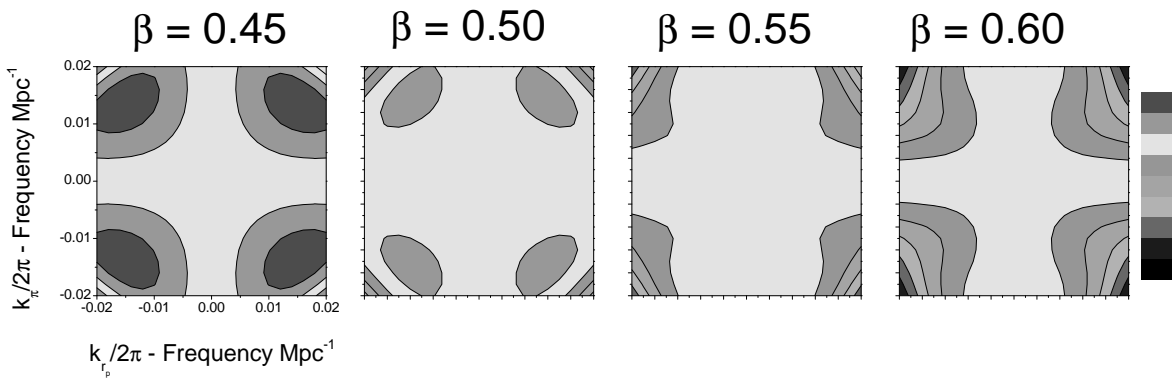} \caption{\it The difference between four
sample values of $\beta$ and the underlying power derived from the signal along
the $k_{r_p}$ axis. Here a much smaller range in $\beta$ was used in order to
show the sensitivity of the method. The inversion of the differences is due to
the different phase of under and over-corrections.}
\end{figure}

Figure 6 shows the `chi-square' values for the appropriate $\beta$ using four
cosmologies and five simulations. This chi-square is simply the sum of the
squares of the difference functions shown in Figure 5, normalized by dividing
each by its minimum value for ease of comparison. The function was calculated
inside a central ring of radius of $k/2\pi = 0.02$, corresponding to
wavelengths greater than or equal to 50 mpc. The sampling resolution of the
difference function is $\Delta k/2\pi = 0.002$.

\begin{figure}[h]
\vbox to2.25in{\rule{0pt}{3.5in}} \includegraphics{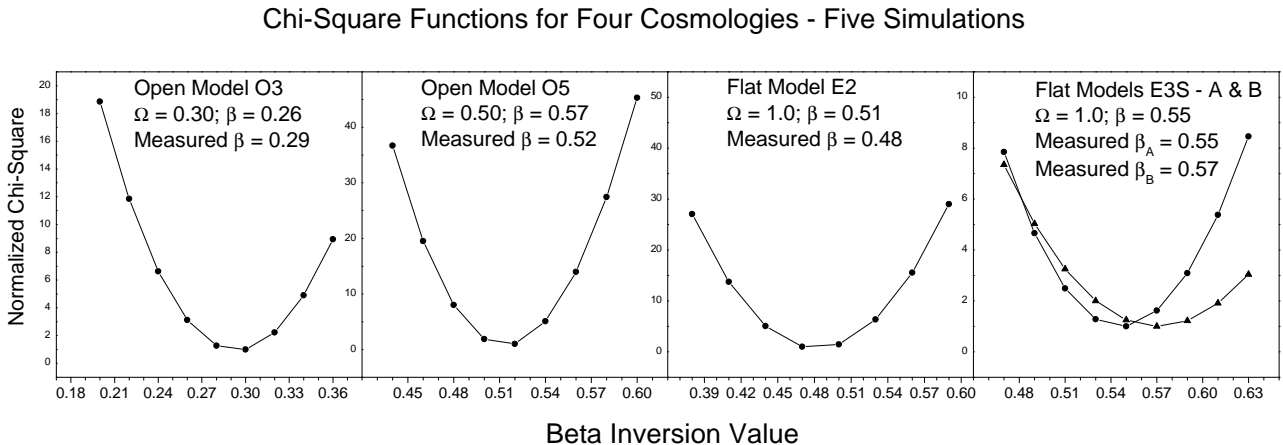} \caption{\it Chi-square functions for the
method using CDM simulations with $\beta=0.26$ ($\Omega_m=0.3$), $\beta=0.57$
($\Omega_m=0.5$), $\beta=0.51$ ($\Omega_m=1.0$), and $\beta=0.54$
($\Omega_m=1.0$). These include COBE, structure normalized, and mixed models.
The uncertainty in $\beta$ across these results is $\pm0.031$. For more detail
on the simulations see the text and Cole \etal (1999).}
\end{figure}

The results include a COBE normalized open $\beta=0.26$ ($\Omega_m=0.3$) model
and a COBE normalized open $\beta=0.57$ ($\Omega_m=0.55$) model, a mixed COBE
and structure normalized flat $\beta=0.51$ ($\Omega_m=1.0$) model, and two
independent structure normalized flat $\beta=0.54$ ($\Omega_m=1.0$) models.
Across these models, $\beta$ has been measured to an uncertainty of $\pm0.031$.
The errors in the estimates are not systematically high or low across the
simulations. This compares favorably to the recent Two-Degree Field result of
$\beta=0.43\pm0.07$ (see Peacock \etal 2001), where the uncertainties are over
a factor of two higher.

\subsection{Difficulties in Applying the Method to Current Survey Data}

Although in theory this method is very straightforward, in practice the
idiosyncracies of real survey data present other challenges. Fortunately, these
difficulties will disappear once the surveys become more contiguous, as they
result from the complexity of current survey geometries.

For the technique to be applied successfully, it is necessary to be able to
break up the survey into thin slices of fairly contiguous data, which do not
include a large number of voids or holes. If there exist a large number of
edges in the data due to a complex angular selection function, then the
undistorted signal will leak out across the edges and into the geometrical
voids in the survey when the data is inverted. This signal is then removed from
the data when the geometric angular selection function cuts are made. This
difficulty is not unexpected when considering that most of the distortion
signal comes from scales above 30 \Mpc, and if the data is not contiguous on
these scales, then the signal will be lost. To improve the method, experiments
are underway to efficiently expand the method to three dimensions to circumvent
this problem.

\section{Conclusion}

A method for determining the value of $\beta=\Omega_m^{0.6}/b$ based upon a
Fourier inversion of distortions in the redshift space density field has been
presented. In contrast to many other measures, this technique does not rely on
a complex modeling of the expected distortions in the redshift space
correlation function or power spectrum. Instead by inverting the redshift space
density distortions as parameterized by $\beta$, it is possible to fit for a
symmetric signal in the redshift space power spectrum. This technique is not
dependent on the small angle/plane-parallel approximation and can make full use
of large redshift survey data. The only present difficultly with this method is
that it does depend on data that is fairly contiguous on large scales. Such
data should become available from large redshift surveys within the next few
years. It has been tested using simulations with four different cosmologies and
returns the value of $\beta$ to $\pm0.031$, greater than a factor of two
improvement over existing techniques. Presently, it has been tested in
two-dimensions using thin slices of data to limit the computational burden, but
can easily be expanded to three-dimensions.

\section{Acknowledgements}

The authors would like to thank S. Cole, S. Hatton, D.H. Weinberg, and C. S.
Frenk (1999) for the use of their simulation data, which they have made easily
accessible to the astronomical community. S. Landy acknowledges support from
the Jeffress Memorial Trust and NSF Grant AST 99-00835. A. Szalay acknowledges
support from ...

\pagebreak

\end{document}